\documentclass[conference]{IEEEtran}
\IEEEoverridecommandlockouts
\usepackage{cite}
\usepackage{amsmath,amssymb,amsfonts}
\usepackage{algorithm}
\usepackage[noend]{algpseudocode}

\usepackage[utf8]{inputenc}
\usepackage{graphicx}
\usepackage{textcomp}
\usepackage{xcolor}
\def\BibTeX{{\rm B\kern-.05em{\sc i\kern-.025em b}\kern-.08em
    T\kern-.1667em\lower.7ex\hbox{E}\kern-.125emX}}
    
\usepackage{stfloats}
    
\begin{document}

\title{Performance Optimization on Model Synchronization in Parallel Stochastic Gradient Descent Based SVM \\
}

\author{\IEEEauthorblockN{1\textsuperscript{st} Vibhatha Abeykoon}
\IEEEauthorblockA{\textit{Intelligent Systems Engineering} \\
\textit{Indiana University}\\
Bloomington, USA \\
vlabeyko@iu.edu}
\and
\IEEEauthorblockN{2\textsuperscript{nd} Geoffrey Fox}
\IEEEauthorblockA{\textit{Intelligent Systems Engineering} \\
\textit{Indiana University}\\
Bloomington, USA \\
gcf@iu.edu}
\and
\IEEEauthorblockN{3\textsuperscript{rd} Minje Kim}
\IEEEauthorblockA{\textit{Intelligent Systems Engineering} \\
\textit{Indiana University}\\
Bloomington, USA \\
minje@iu.edu }
}

\maketitle

\newcommand{\norm}[1]{\left\lVert#1\right\rVert}

\begin{abstract}
Understanding the bottlenecks in implementing stochastic gradient descent (SGD)-based distributed support vector machines (SVM) algorithm is important in training larger data sets. The communication time to do the model synchronization across the parallel processes is the main bottleneck that causes inefficiency in the training process. The model synchronization is directly affected by the mini-batch size of data processed before the global synchronization. In producing an efficient distributed model, the communication time in training model synchronization has to be as minimum as possible while retaining a high testing accuracy. The effect from model synchronization frequency over the convergence of the algorithm and accuracy of the generated model must be well understood to design an efficient distributed model.  In this research, we identify the bottlenecks in model synchronization in parallel stochastic gradient descent (PSGD)-based SVM algorithm with respect to the training model synchronization frequency (MSF). Our research shows that by optimizing the MSF in the data sets that we used, a reduction of 98\% in communication time can be gained (16x - 24x speed up)  with respect to high-frequency model synchronization. The training model optimization discussed in this paper guarantees a higher accuracy than the sequential algorithm along with faster convergence.

\end{abstract}



\section{Introduction}\label{s:intro}

Support vector machines (SVM) algorithm is an important classification algorithm in the supervised machine learning domain. In training SVM for larger data sets, the most important thing is to identify the bottlenecks in training. The main reason is that the time to train an SVM is computationally higher when it comes to dealing with high volume data with higher dimensions. The distributed version of SVM is an effective solution to this problem. In scaling distributed support vector machines algorithm, the most important thing is to determine the bottlenecks in scaling the algorithm. The number of processes and limitation of resources in the distributed environment \cite{scaling-svm} is vital to determine optimum performance. In scaling the algorithm across the cluster resources, there are two types of overheads that have to be dealt with. The major challenge is avoiding the communication overhead in synchronizing distributed models which causes a lag in performance. The next challenge is to identify the core algorithm used in the SVM to minimize the computation overhead. In referring to the computation overhead, there are many versions of SVM algorithm which has provided various optimization to improve the performance in the sequential algorithm \cite{svm-vapnik}, \cite{dc-svm}, \cite{libsvm}, \cite{platt}, \cite{pegasos}. In parallel, fast SVM algorithms will reduce the computational load in the nodes of the cluster. This fact, however, will make communication overhead more pronounced. Consequently, the next natural step is to optimize communication. \cite{psvm}, \cite{psvm1}, \cite{p-packsvm}. In order to provide a faster parallel training model, the distributed algorithm must be communication-efficient and computation-efficient. The communication model decides the final value of the globally-synchronized model. So, the configurations in the communication model have to be optimized to obtain higher accuracy. The training model synchronization across the distributed nodes is a very important fact to gain higher accuracy and efficiency in training. In this research, we thoroughly look into the communication overhead caused by the model synchronization against the frequency of synchronization. In an optimum and efficient distributed model, the communication cost should be relatively lesser than the computation cost. With the faster execution, maintaining a highly accurate model has to be well understood. Scaling the distributed algorithm by maintaining the output quality as same as the sequential algorithm is vital to guarantee the consistency of the scaled algorithm. Throughout this research, we analyze the effects of faster execution over the accuracy of the trained model. And also we analyze how to minimize the bottlenecks in scaling the SVM algorithm in distributed environments. In this paper, we discuss how to guarantee higher testing accuracy and faster convergence by optimizing model synchronization along with minimizing communication overhead caused by frequent model synchronization in the distributed paradigm. In section \ref{s:relatedwork} we discuss the related work done on SVM, in section \ref{s:background}, the mathematical aspects of the SVM training model is discussed. In section \ref{s:methodlogy}, the nature of the traditional sequential algorithm and the effect by the frequent model synchronization in the parallel algorithm is discussed along with simulating that effect on the sequential version of the parallel algorithm. In section \ref{s:experiments}, the conducted experiments and the results are explained with respect to the methodologies discussed in section \ref{s:methodlogy}. The conclusions and future work of the current research are discussed in section \ref{s:conclusion}.
\section{Related Work}\label{s:relatedwork}

Support Vector Machines (SVM) by Cortes and Vapnik \cite{svm-vapnik} can be considered as one of the earliest methodologies used in the supervised learning-based classification. There are couple of sequential implementations like DC-SVM \cite{dc-svm}, LibSVM \cite{libsvm} and Sequential Minimal Optimization (SMO) \cite{platt}, \cite{smo-like}, \cite{ssmo}, \cite{ssmoimp} which can be considered as most prominent sequential implementations to solve the SVM problem. Building an SVM model becomes computationally expensive depending on the number of data points and the dimension of a data point in the data set. For a data set having few hundreds of Mega Bytes can cause memory bound issues when the algorithm has to compute a kernel matrix of size $n \times n$ where $n$ is the number of data points in the data set. To overcome this problem there have been many studies done considering random samples via bootstrap techniques \cite{svm-ensemble}, described in SVM ensemble. In these studies, the performance improvement on nature of execution for very large data sets has not been elaborated. In LibSVM, DC-SVM and most of the SVM-based implementations, the core algorithm used is the computationally expensive SMO algorithm. The parallel implementations, done on SMO-based SVM by Keerthi et al in \cite{smo-parallel} can be recognized as one of the earliest work on this problem. But the SMO itself is a computationally intensive model due to the high overhead in optimizing Lagrangian multipliers in an iterative way. Apart from parallel SMO, there have been matrix approximation methods that have been used to work on the memory-based overhead in the traditional SMO algorithm \cite{psvm}, \cite{psvm1}. The matrix factorization methods and SMO-based algorithms are still computationally intensive and it doesn't provide a pleasingly parallel model. 

In recent research, to avoid this problem involved with model parallelism overhead, gradient descent-based optimization \cite{psgd} has been widely used. The main reason for the overhead in model parallelism in SMO and matrix-based approaches is due to the high computation and communication overhead in solving the quadratic objective function using a linear equation system. In this regard, SGD-based approaches are an alternative solution because the computation of stochastic gradient step is much faster than solving a set of linear equations. Instead of solving a linear equation system, the problem can be solved by minimizing the objective function using a traditional SGD-based approach and estimating the weights that satisfy the minimal objective function. The SGD-based approaches have been widely used in pPackSVM \cite{p-packsvm} and fast feature extracting SVM approaches \cite{ff-svm}. Pegasos \cite{pegasos} is another prominent SGD-based SVM optimization done with an adaptive decreasing learning rate, which provides a guaranteed convergence in a lesser number of epochs.

\section{Background} \label{s:background}
In our research we focus on linear kernel-based binary classification on three different data sets. The Epsilon \cite{ds-epsilon} dataset contains 400,000 samples with 2,000 features; Ijcnn1 \cite{ds-ijcnn1} dataset contains 35,000 samples with 22 features; Webspam \cite{ds-webspam} contains 350,000 samples with 254 features. In referring to the mathematical background associated with SGD-based SVM, in the sample space of $S$ with $n$ samples, ${x_{i}}$ refers to a $d$-dimensional feature vector and ${y_{i}}$ is the label of the ${i^{th}}$ data point as shown in (\ref{eq:1}).

\begin{align}
\nonumber    S &= \{x_{i}, y_{i}\} \\
\label{eq:1}\text{where} & ~i = [1,2,3,...,n], ~x_{i} \in R^{d}, ~y_{i} \in [+1, -1]
\end{align}

In the SGD approach the objective is to minimize the objective function in (\ref{eq:2}) with the constraint on the optimization defined in (\ref{eq:3}):

\begin{align}
    \label{eq:2}J^{t} &= \min_{w \in R^{d}} \frac{1}{2} \norm{w}^2 + C \sum_{x,y \in S} g(w;(x,y))\\
    \label{eq:3}    g(w;(x,y)) &= \max( 0, 1 - y \langle w|x \rangle )
\end{align}

Depending on the value of the constraint function, the weight update will be done as in (\ref{eq:7}) with the learning rate $\alpha$ by  considering (\ref{eq:5}) as the derivative depending on the value obtained for the expression in (\ref{eq:4}). $C$ in (\ref{eq:2}) refers to a tuning hyper parameter. $w$ in (\ref{eq:2}) refers to the weight vector. 

\begin{align}
\label{eq:7}    w &= w - \alpha \nabla J^{t}, ~ \alpha = \frac{1}{1+t}\\
\label{eq:5} \nabla J^{t} &= \left\{\begin{array}{cl} w & \text{if  }\max( 0, 1 - y \langle w|x \rangle ) = 0\\
w - C x_{i} y_{i} & \text{Otherwise}\end{array}\right.\\
\label{eq:4}    y \langle w | x \rangle &= y_{i} {w}^\top x_{i}
\end{align}

In the experiments conducted in this paper, we use a learning rate of $\alpha$ which is decaying with the epoch number $t$. In the current research, the communication overhead in frequent-model-synchronization has not been thoroughly discussed with respect to the convergence of the algorithm on lower objective function value and higher cross-validation accuracy. In this paper, we analyze how the MSF affects the faster convergence and faster execution of the distributed SGD-based SVM algorithm.  

\section{Methodology}\label{s:methodlogy}

Our objective is to analyze the effect of model synchronization on faster convergence and to see how model synchronization communication overhead can be optimized to run the training model in an efficient way. In this regard, we modified the original sequential algorithm to get the same effect caused by the parallel model synchronizing algorithm to verify the accuracy of the distributed model we built. The model synchronization resembles synchronizing the training weight vector in all parallel machines by averaging over the sum of the local weight vector in each machine. Model synchronization in the distributed mode considers each machine in the system as a single block of a sequential algorithm, where each machine updates the model per each data point and do a model synchronization after each machine has calculated the corresponding model. This is the atomic-level model synchronization that can take place in the frequent model synchronization. It is clear that there can be a model synchronization overhead caused by frequent synchronization due to inter-process communication and this effect will be addressed in section \ref{s:experiments}. In this paper, we refer to a term called model synchronization frequency (MSF) which refers to the number of data points used to calculate the model before synchronizing it with other processes in the distributed training mode. Note that if data points used for model synchronization is unity ($=1$), it is considered as a higher MSF as we will be synchronizing the models after each data point in each process is done with calculating the model. If the data points used per synchronization is a large number $L$, it means there will be a model synchronization happening after calculating the weight for $L$ data points in each process which implies that the MSF is low. First, we focus on determining the accuracy of the distributed model synchronizing algorithm that we introduce with respect to the sequential version of the distributed computation model without communication implementation. Then we focus on implementing the distributed version with configurable MSF value to observe the convergence of the algorithm with respect to higher cross-validation accuracy (without overfitting) and lower value of the objective function.

\subsection{Standard Sequential Algorithm}\label{s:s:seq_a}

In the standard sequential version of SGD-based SVM in algorithm \ref{alg:ss}, the weights are initialized with a Gaussian distribution and the training process is done for $T$ iterations. The value for $T$ is decided by prior experiments, where both cross-validation accuracy and the value of the objective function are considered in a such a way that both values come to a stage where their oscillations are in a minimum level. The reason for picking a constant $T$ is for the convenience in timing comparisons and to see how each tuning parameter affects the convergence in the experiments conducted with methods in \ref{s:s:model-seq-synch} and \ref{s:s:model-par-synch}. $T=T_{i}$ is the iterations towards convergence point in ${i^{th}}$ data set. And this value is different from dataset to dataset, but in the experiments we observe the variation of accuracy and value of the objective function even after convergence to show case the consistency of the training after the convergence.

\begin{algorithm}[ht]
\caption{Sequential Stochastic Gradient Descent SVM}
\begin{algorithmic}[1]  
\State \textbf{$INPUT  : [x,y] \in S ,w \in R^{d}$}
\State \textbf{$OUTPUT : w \in R^{d}$}
\Procedure{SGD}{$S,w$}  
\For {$t = 0 $ to $T$}
\For {$i = 0 $ to $n$}
\If{$(g(w;(x,y))==0)$}
\State  $\nabla J^{t} = w$
\Else
\State  $\nabla J^{t} = w - C x_{i} y_{i}$
\EndIf
\State $w = w - \alpha \nabla J^{t}$
\EndFor
\EndFor
\Return $w$
\EndProcedure  
\end{algorithmic}  
\label{alg:ss}
\end{algorithm}


\subsection{Sequential Replica of Distributed Model Synchronizing Algorithm}\label{s:s:model-seq-synch}

The sequential replica of distributed SGD-based SVM model synchronizing algorithm (SRDMS-SGD-SVM) in algorithm \ref{alg:sms} is important because we are altering the order of updating the weights in the parallel mode with respect to the standard SGD-based SVM algorithm. The objective is to show case how the distributed algorithm can be modelled in a sequential manner. By this we only intend to implement the computation model of the distributed algorithm in a sequential manner. Hence, it is important to see the effect of the sequential version of the same algorithm is as same as the parallel version of the algorithm. This way we can analyze the accuracy of the implemented algorithm. In the sequential version of the model synchronizing algorithm, we shuffle the data before creating the small blocks of data resembled as $S_k$ in algorithm \ref{alg:sms}. Each block resembles the union of a chunk of data which will be processed in the parallel mode in each process before doing the model synchronization. The block size of unity resembles the standard SGD-based algorithm. For each data point in each block, the weight vector is the same unlike in the standard SGD algorithm. In parallel mode, if the parallelism is $k$, and the model synchronization frequency is one, the initial weights of each element processed in each processor is same, implying the sequential version of the algorithm must have the same quality, that is the intuitive idea behind the different block sizes chosen in the sequential algorithm. But, in the sequential algorithm, the model synchronization does not mean that it is doing any communication over the network. Instead, it merely represents the idea that each block will have the same initial weight for each element in a given block. When updating the weight vector at the end of each model synchronization, we calculate the average of the weights calculated per each data point in the block to determine the final weight. Another way of doing this is by considering a weighted model synchronization based on pre-determined parameters. There pre-determined parameters are mostly depending on the data partitioning mechanism used. In this research we used a random data partitioning technique. So we only consider about a non-weighted model synchronization.  

\begin{algorithm}[ht]
\caption{Sequential Replica of Distributed Model Synchronizing Algorithm}
\begin{algorithmic}[1]  
\State \textbf{$INPUT  : S \in [S_{1}, S_{2} ..., S_{b}], w \in R^{d}$}
\State \textbf{$OUTPUT : w \in R^{d}$}
\Procedure{SRDMS}{$S,w$}
\For {$k=1$ to $b$}
\Procedure{SGD}{$S_{k},w$}  
\EndProcedure 
\EndFor
\EndProcedure
\Return $w$
 
\end{algorithmic}  
\label{alg:sms}
\end{algorithm}

\subsection{Distributed Model Synchronizing Algorithm}\label{s:s:model-par-synch}

In the model synchronizing distributed SGD-based SVM in algorithm \ref{alg:dms}, the training data set is loaded in a way that equal amount of data is loaded to each machine in order to balance the load among processes. Load balancing is compulsory in order to reduce the process waiting time caused when a subset of processes or a process takes a longer time to complete the computation than the rest. Each machine loads the data and shuffles before the training process. The shuffled data is then processed as blocks in the training process. When the block size defined in the training process is $s_{b}$, each machine will run the sequential SRDMS SVM algorithm on $s_{b}$ number of data points in each process (a block is processed) and the model synchronization is done after processing each block with $s_{b}$ data points in each of the processes. For each block, there will be its own local model which communicates with each of the machines using \texttt{MPI\_AllReduce}. The average of this is used as the global model after each model synchronization. The global model then becomes the initial weights for the next block. The $T$ training iterations are conducted to reach convergence. 

\begin{algorithm}[ht]
\caption{Distributed Model Synchronizing Algorithm (DMS)}
\begin{algorithmic}[1]  
\State \textbf{$INPUT  : S \in [S_{1}, S_{2} ..., S_{b}], w \in R^{d}, |(j-i)|*K = b $}
\State \textbf{$OUTPUT : w \in R^{d}$}
\Procedure{DMSSGD}{$S,w$}
\State \textbf{In Parallel in K Machines $[S_1,...,S_b] \subset S$ }
\State \textbf{$w_{local}=w$}
\For {$m=i$ to $j$}
\Procedure{SRDMS}{$S_{m},w_{local}$}  
\EndProcedure 
\State \textbf{$w_{global} = \texttt{MPI\_AllReduce}(w_{local})$}
\State \textbf{$w = w_{global}/K$} 
\EndFor
\EndProcedure
\Return $w$
\end{algorithmic}  
\label{alg:dms}
\end{algorithm}


\section{Experiments}\label{s:experiments}

In the experiments, we focused on three main sections to analyze how to optimize the training process to gain faster execution with higher accuracy. In the first set of experiments, we observe how the model synchronization variation affects the convergence of the sequential algorithm and analyze how that information can be used to optimize the distributed version of the algorithm. In the second set of experiments, we change the model synchronization frequency (MSF) and parallelism to observe how convergence can be obtained. In the third set of experiments, we analyze the execution time variation with respect to the variation of MSF for different parallelisms. From these experiments, we analyze how an optimized training model can be obtained to guarantee faster execution and higher testing accuracy \cite{iu-psgdsvmc}. Ijcnn1, Webspam and Epsilon are the datasets which are used in the experiments. In Table \ref{tb:dataset}, the nature of the data sets is shown with respect to training, cross-validation, testing, feature size and sparsity percentage. The experiments were conducted in Intel(R) Xeon(R) distributed cluster hosted in Future Systems. For the experiments, we use single node core level parallelism for smaller data sets and for the largest data set we use node-level parallelism with Infiniband support in OpenMPI. Throughout the experiments we kept the hyper-parameter $C=1$ in (\ref{eq:2}) and the learning rate as an adaptive diminishing function as in (\ref{eq:7}).

\begin{table*}[t]
\caption{Datasets}
\begin{center}
\begin{tabular}{|l|l|l|l|l|l|}
\hline
\textbf{DataSet} & \textbf{Training Data (60\%,80\%)} & \textbf{Cross-Validation Data (60\%,80\%)} & \textbf{Testing Data (60\%,80\%)} & \textbf{Sparsity} & \textbf{Features} \\ \hline
Ijcnn1 & 21000 ,28000 & 7000,3500 & 7000,3500 & 40.91 & 22 \\ \hline
Webspam & 210000,280000 & 70000,35000 & 70000,35000 & 99.9 & 254 \\ \hline
Epsilon & 240000,320000 & 80000,40000 & 80000,40000 & 44.9 & 2000 \\ \hline
\end{tabular}
\end{center}
\label{tb:dataset}
\end{table*}

\subsection{Effect of Block Size on Sequential Algorithm Convergence}

The objective of these sequential experiments is to analyze how the variation of model synchronization or block size affects cross-validation accuracy.  In sequential mode, the model synchronization resembles the computational equivalence of SRDSM in algorithm \ref{alg:sms} to DMS in algorithm \ref{alg:dms}.



\begin{figure}[!t]
\includegraphics[width=0.45\textwidth]{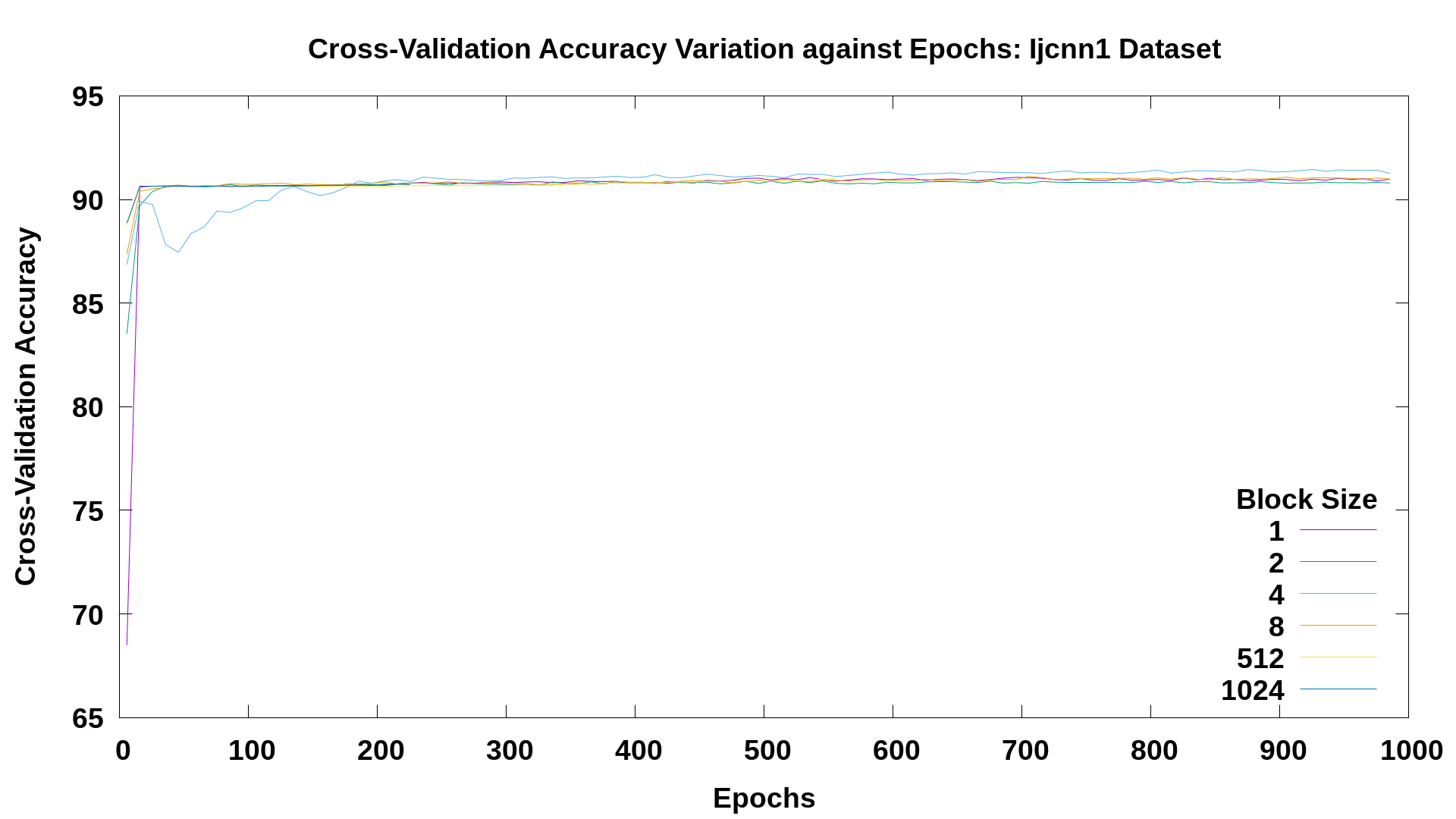}
\caption{Cross Validation Accuracy Variation in Sequential Algorithm for Ijcnn1 Dataset : Block Size = [1,2,4,8,512,1024] }
\label{fig:seq-msf-ijcnn-1}
\end{figure}

\begin{figure}[!t]
\includegraphics[width=0.45\textwidth]{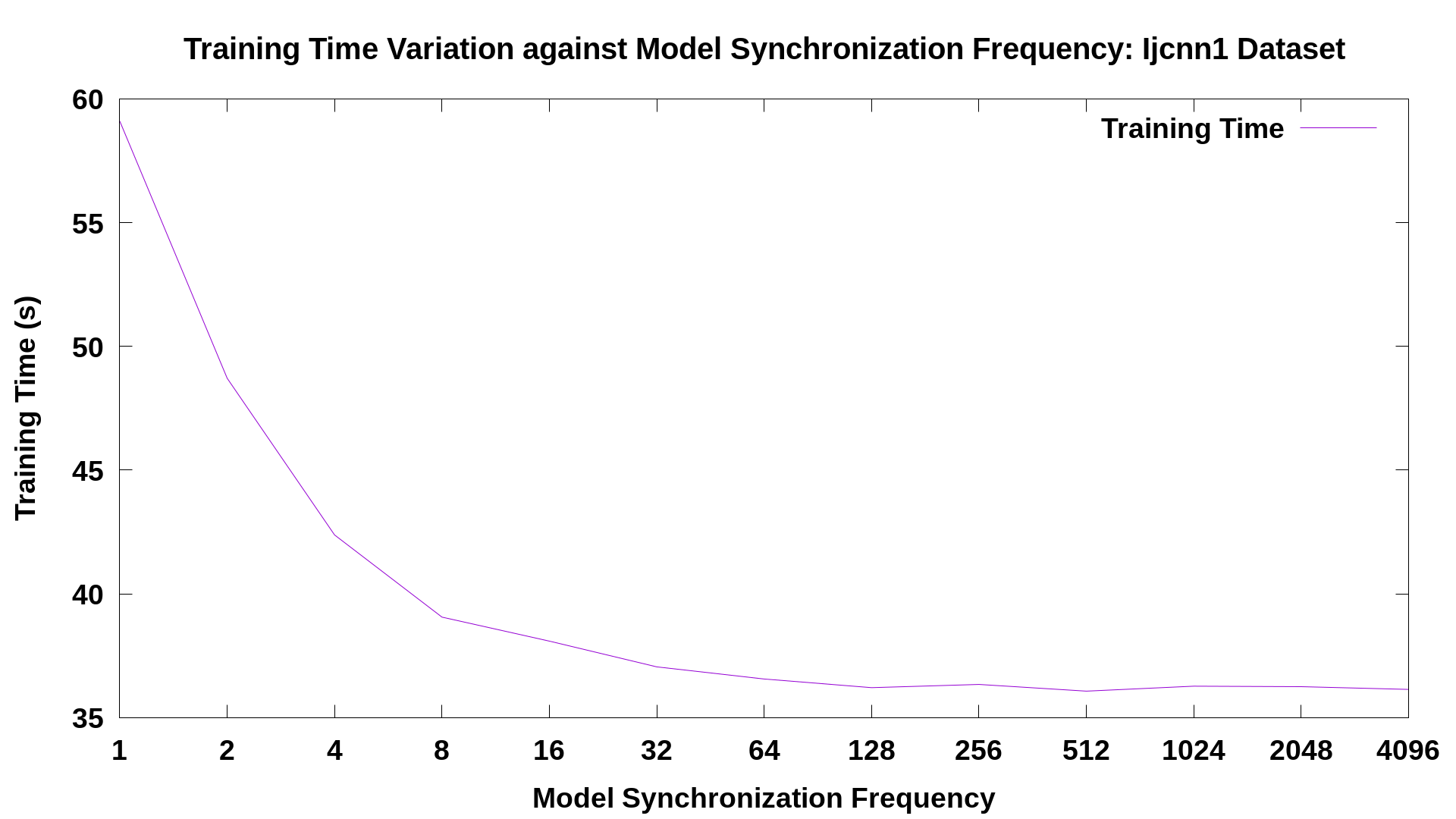}
\caption{Training Time Variation with Variable Block Sizes on Ijcnn1 Dataset  }
\label{fig:seq-msf-ijcnn-2}
\end{figure}



\begin{figure}[t]
\includegraphics[width=0.45\textwidth]{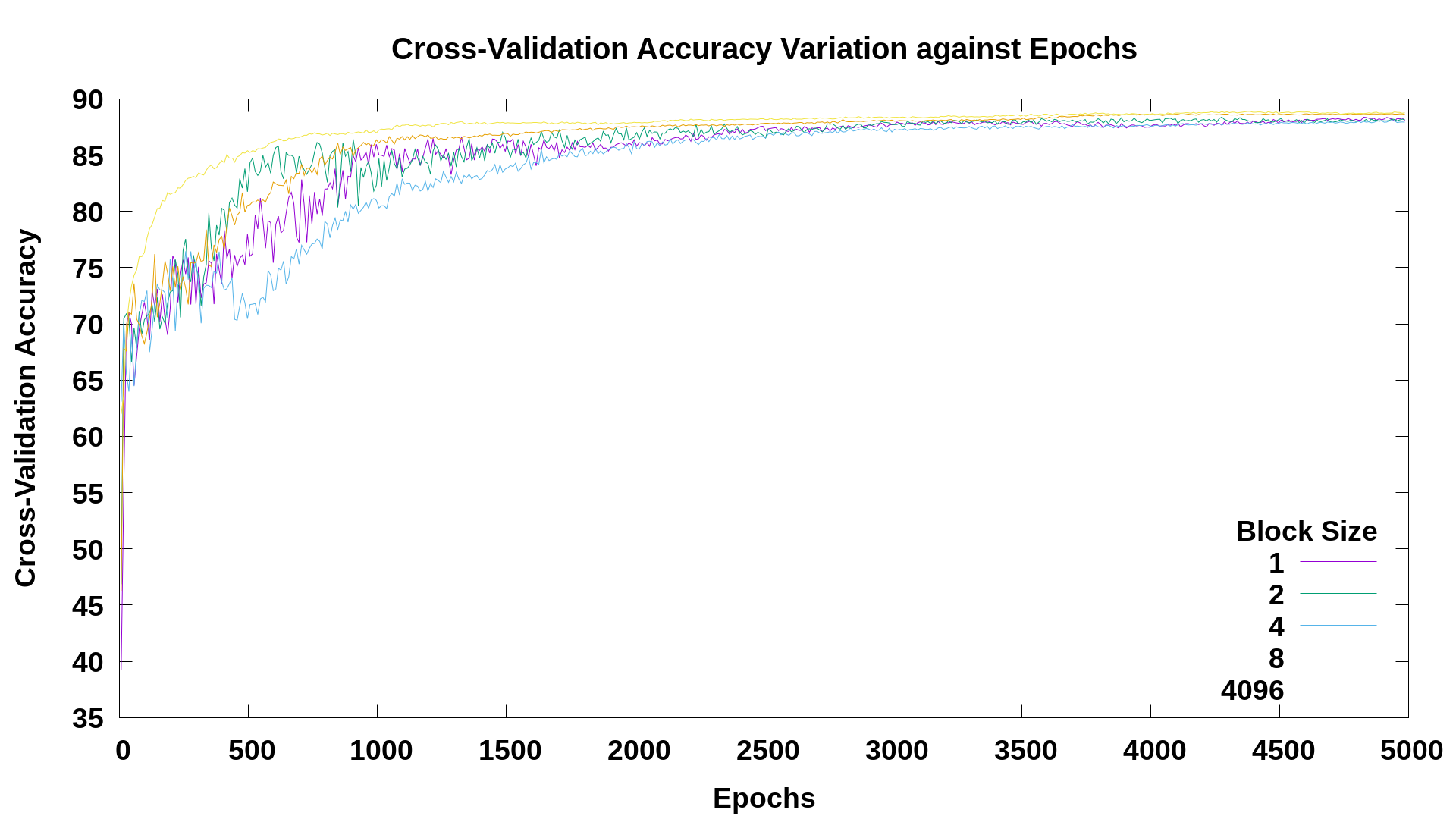}
\caption{Cross Validation Accuracy Variation in Sequential Algorithm for Webspam Dataset : Block Size = [1,2,4,8,4096] }
\label{fig:seq-msf-webspam-1}
\end{figure}

\begin{figure}[t]
\includegraphics[width=0.45\textwidth]{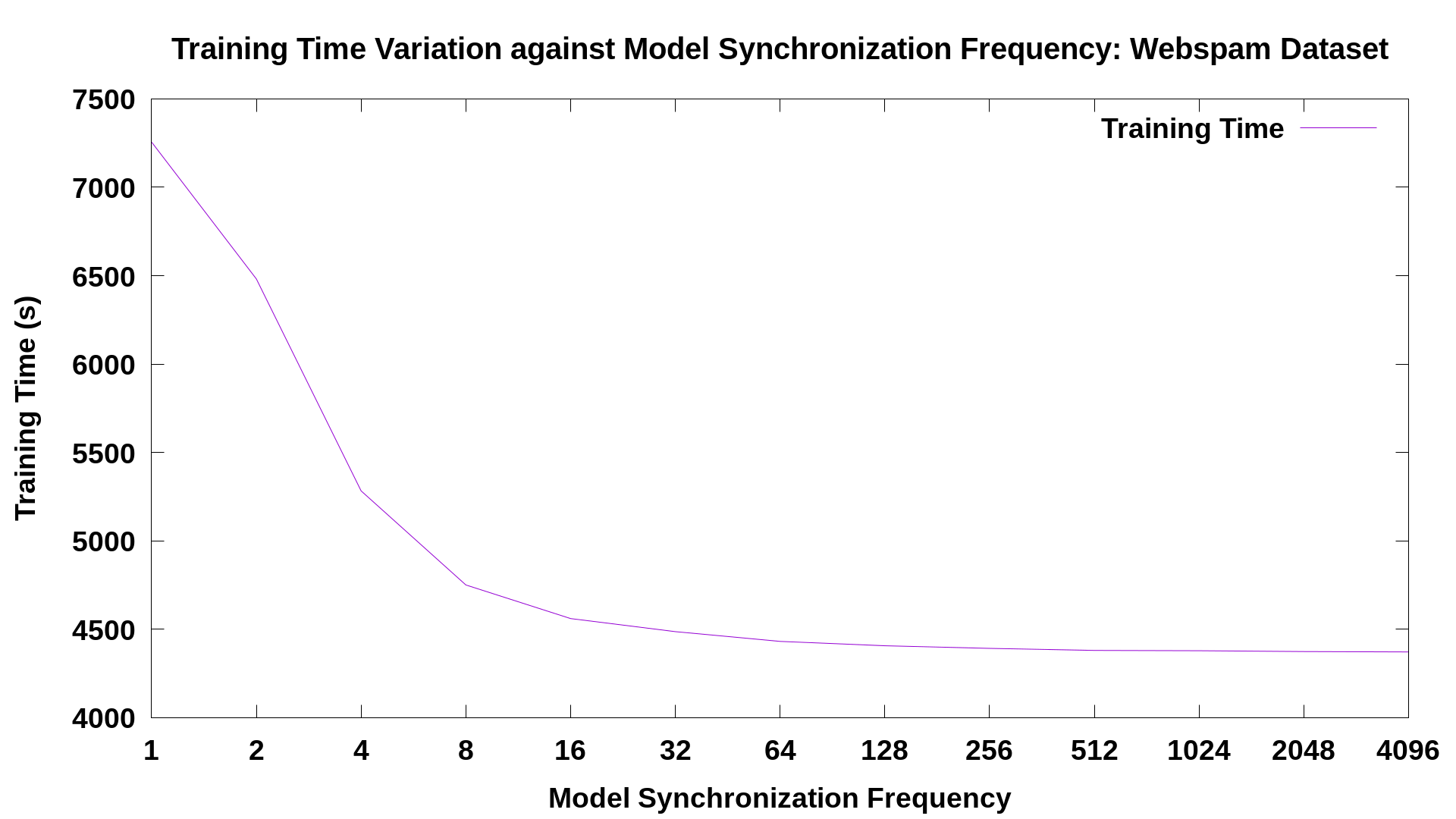}
\caption{Training Time Variation with Variable Block Sizes on Webspam Dataset}
\label{fig:seq-msf-webspam-2}
\end{figure}

In Ijcnn1, we considered the block sizes 1,2,4,8,1024 and block sizes 1,2,4,8,4096 for Webspam dataset to analyze the effect on cross-validation accuracy. Figures \ref{fig:seq-msf-ijcnn-1} and \ref{fig:seq-msf-webspam-1} show experiments on Ijcnn1 and Webspam data sets. The effect from MSF over cross-validation accuracy is approximately negligible in these two data sets. In the experiments, we initialized multiple experiments with unique Gaussian initialization for each experiment. Then we averaged the cross-validation accuracy over multiple experiments with better convergence. From experiments, we learned that the convergence of Ijcnn1 data set is highly sensitive on the initialization unlike Webspam data set. With this experiment setting, we were able to get an accurate conclusion on the cross-validation accuracy variation with respect to variable MSF. Figures \ref{fig:seq-msf-ijcnn-2} and \ref{fig:seq-msf-webspam-2} shows the training time variation with MSF variation. When the MSF is high (lower block size), the training time is much higher and the training time dilutes down and becomes steady after a threshold MSF value. This observation implies that with a higher block size (lower MSF) allows the algorithm to run much faster with the same convergence. The reason for time dilution with higher block size comes with the less overhead caused by average model calculation and cross-validation accuracy calculation as they are executed when the model synchronization is done. When MSF is lower, the frequency of model averaging and cross-validation accuracy calculation is lower and it provides a performance boost. The results gathered from this section, reveals how the block size affects the convergence of the algorithm. Here we used two data sets, Webspam and Ijcnn1 to analyze what happens with the variable block sizes. It is evident from the observations from this section that for SGD-based SVM the block size or mini-batch size provides a less variation on the cross-validation accuracy.

\subsection{Model Synchronization Effect on Parallel and Sequential Algorithm Convergence}

In understanding how the frequent model synchronization can affect the cross-validation accuracy in the sequential and the parallel version, we conducted experiments to see how the cross-validation accuracy behaves in the training period. The main objective of these experiments is to see whether the parallel algorithm and sequential algorithm behaves in a similar way towards convergence. Using these experiments, we can verify the approach we used in the experiments is accurate. The comparison of distributed model synchronization along with the replica of it in sequential mode allows us to show that the developed model is functionally accurate. We used Ijcnn1 and Webspam data sets to see how the distributed and sequential models can provide similar results. All of the results of these experiments can be referred from \cite{iu-psgdsvmc}. From the results obtained in this section, it is evident that the distributed model and sequential model provide similar results ensuring that the distributed model functions accurately. This observation ensures that the designed model is functionally accurate in distributed paradigm. 

\begin{figure}[t]
\centering
\includegraphics[height=0.45\textwidth,width=0.50\textwidth]{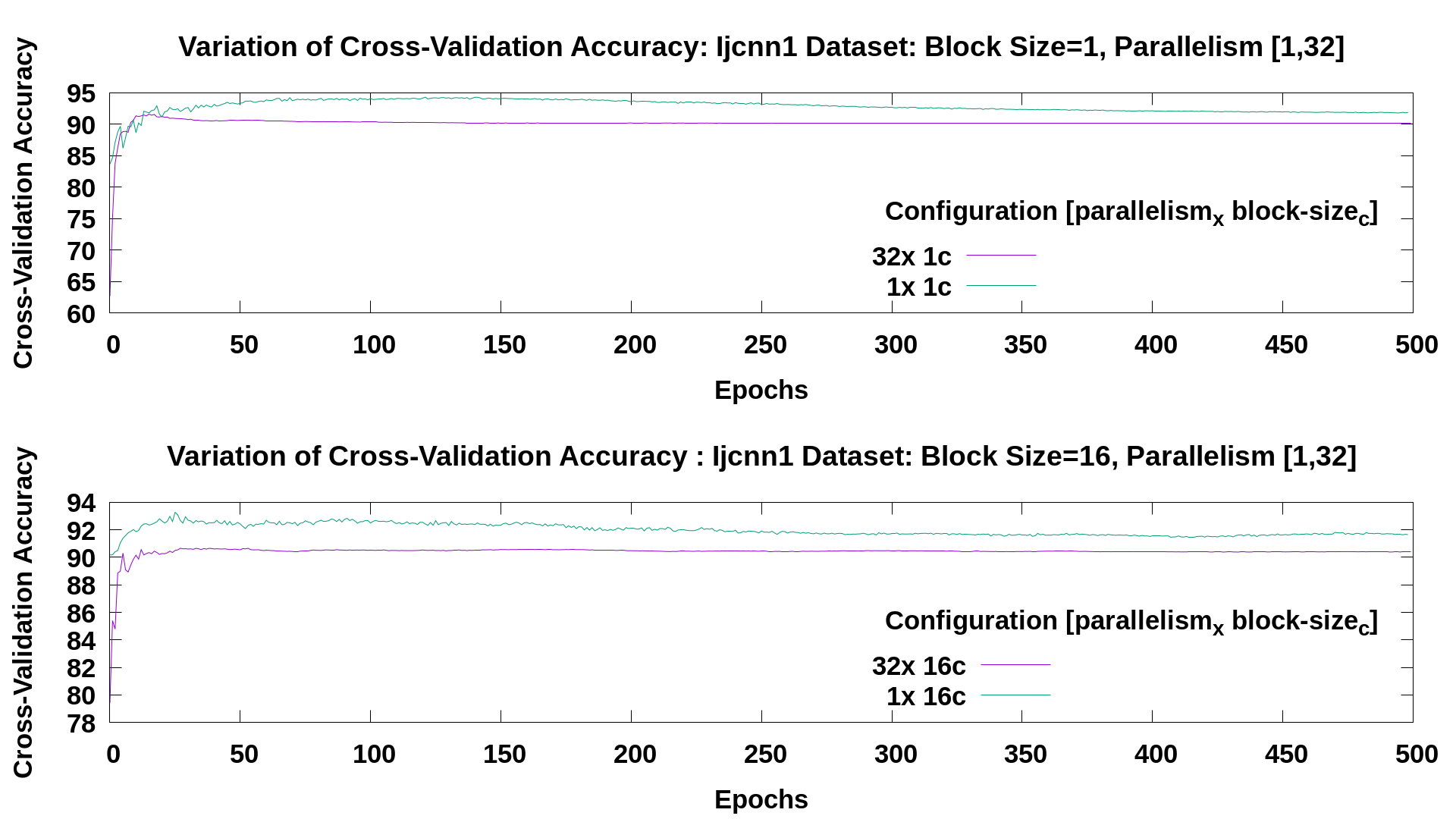}
\caption{Cross Validation Accuracy  Variation against Block Size with Parallelism : Ijcnn1 Dataset}
\label{fig:seq-msfp-ijcnn1-32x-1}
\end{figure}

\begin{figure}[t]
\includegraphics[height=0.45\textwidth,width=0.50\textwidth]{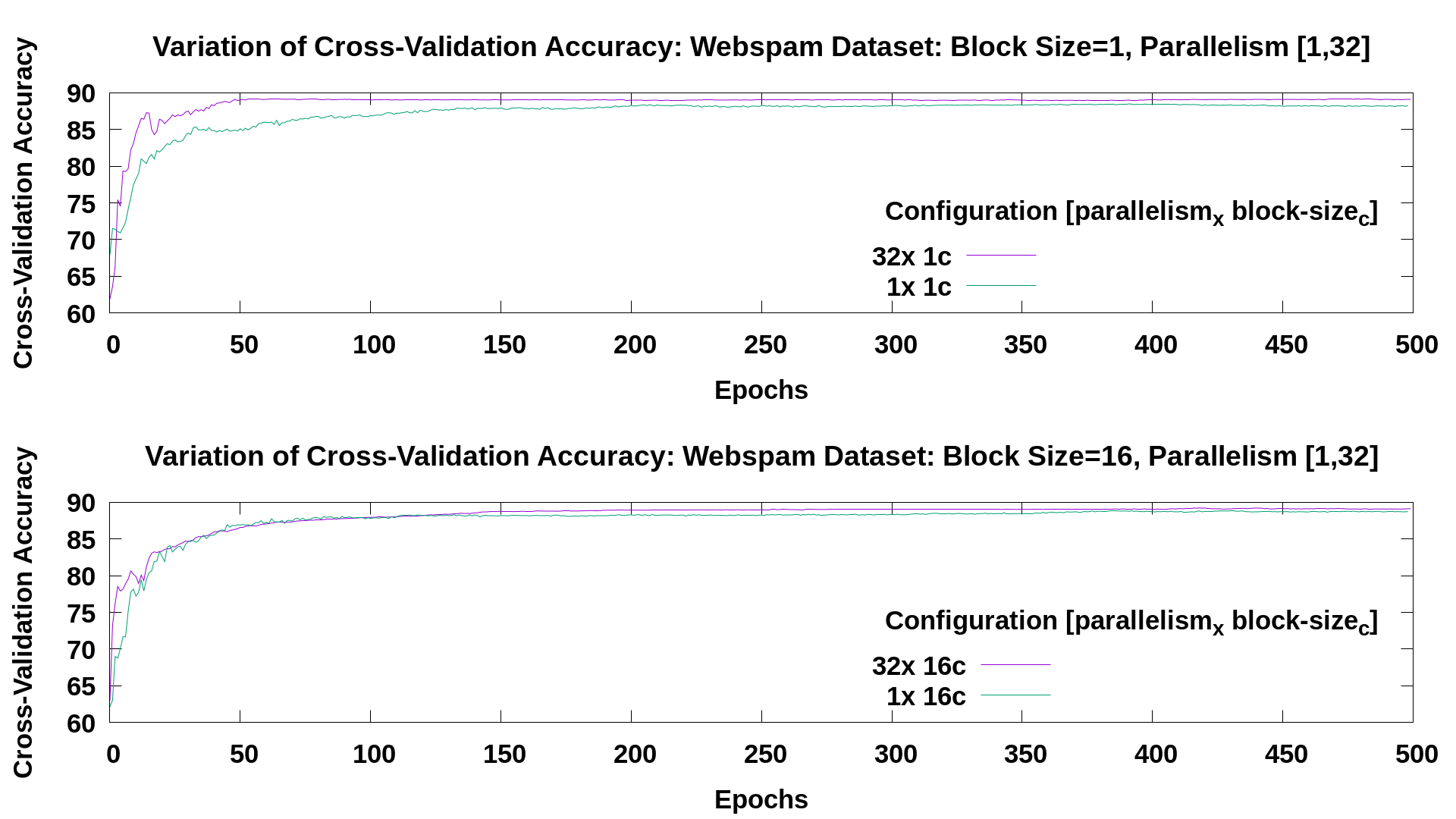}
\caption{Cross Validation Accuracy Variation against Block Size with Parallelism : Webspam Dataset}
\label{fig:seq-msfp-webspam-32x-1}
\end{figure}


\subsection{Distributed Model Synchronizing Experiments}

\subsubsection{Distributed Model Synchronization on Convergence}

In the distributed model synchronizing experiments, we evaluate how the algorithm convergence is affected by the model synchronization frequency along with the variation of the parallelism. We consider three groups of experiments for this, in the first one we consider the high-frequency range where it involves 1-8 block sizes. In the second group, 16-512 block sizes for mid-range frequencies and for the third group, lower frequencies with 512-4096 block sizes were used. The reason behind variable frequency groups comes when we have a data set with limited data size, we can only pick up to a certain set of frequencies. For instance, in Ijcnn1 data set, the total training data points are 28,000 (80\% of data for training) and if we have 32 processes to do the computation, a single process will have only 875 data points, so the maximum block size we can use is 875 and the minimum is 1 and they correspond to lowest MSF and highest MSF respectively. In figures \ref{fig:dms-cost-ijcnn1-32x-1,2,4,512}, \ref{fig:dms-cost-webspam-32x-1,2,4,512} and \ref{fig:dms-cost-epsilon-32x-1,2,4,512}, the parallel experiments on parallelism 32 for variable MSF is shown. It is clear from these experiments that the variation of cross-validation accuracy and value of the objective function remains approximately the same with the denoted MSFs on all three data sets. From these observations, we can understand that frequent model synchronization is not highly vital to obtain a higher cross-validation accuracy or a lower objective function value, irrespective of the nature of the data set. As same as in the sequential experiments, the random Gaussian initialization is vital to identify the best initialization which provides an accurate output. This is the same realization we obtained from the sequential version of the algorithm and by these results, we can verify the parallel model we developed is consistent with the sequential form of the modified SGD-based SVM algorithm. We conducted these experiments for parallelism 2,4,8,16 and 32 and all these results can be seen in \cite{iu-psgdsvmc}.

\begin{figure}[t]
\includegraphics[height=0.45\textwidth,width=0.50\textwidth]{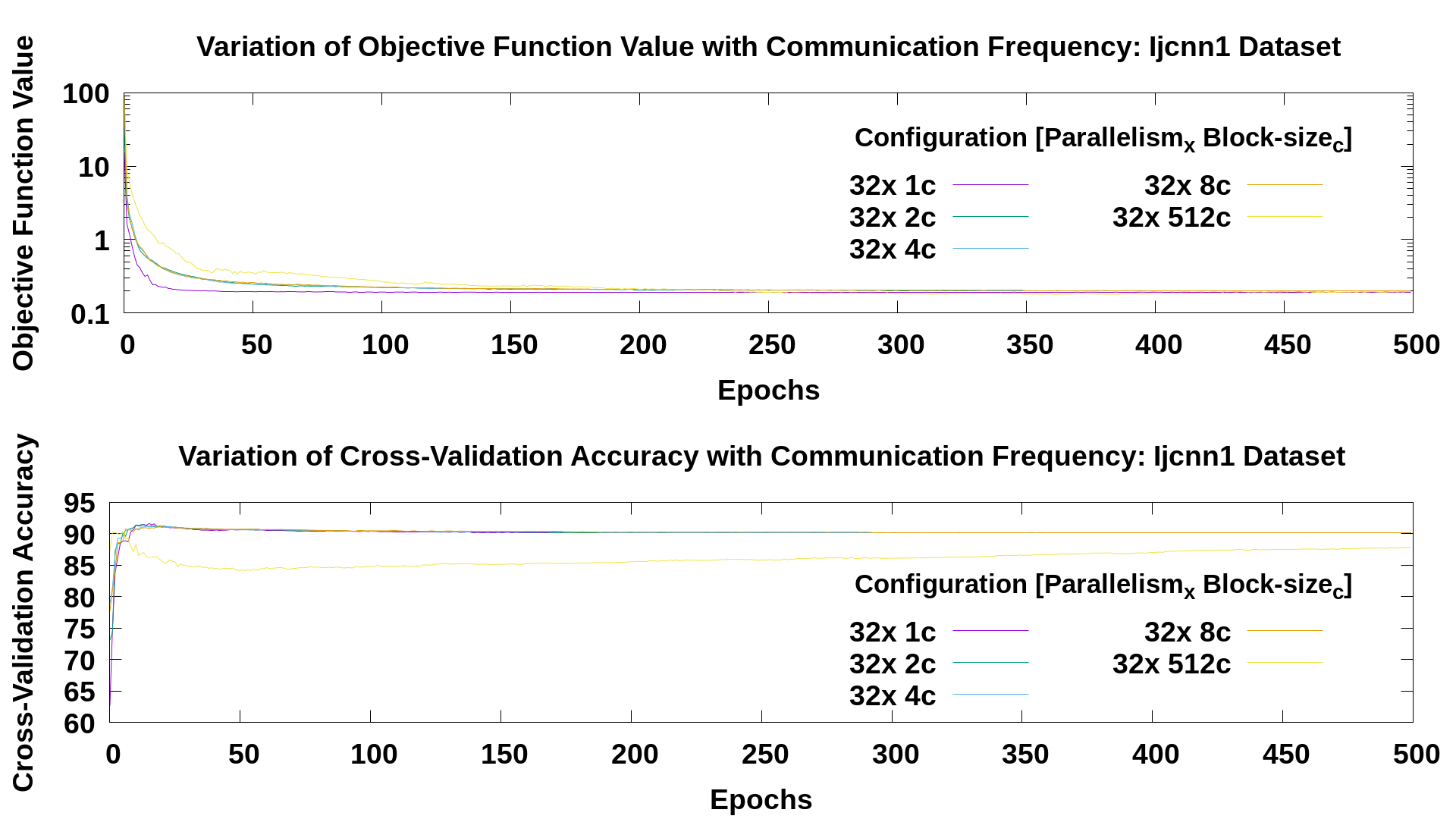}
\caption{Distributed Model Synchronizing Algorithm on Ijcnn1 Dataset for Parallelism [32] and Block Size = [1,2,4,8,512]}
\label{fig:dms-cost-ijcnn1-32x-1,2,4,512}
\end{figure}



\begin{figure}[t]
\includegraphics[height=0.45\textwidth,width=0.50\textwidth]{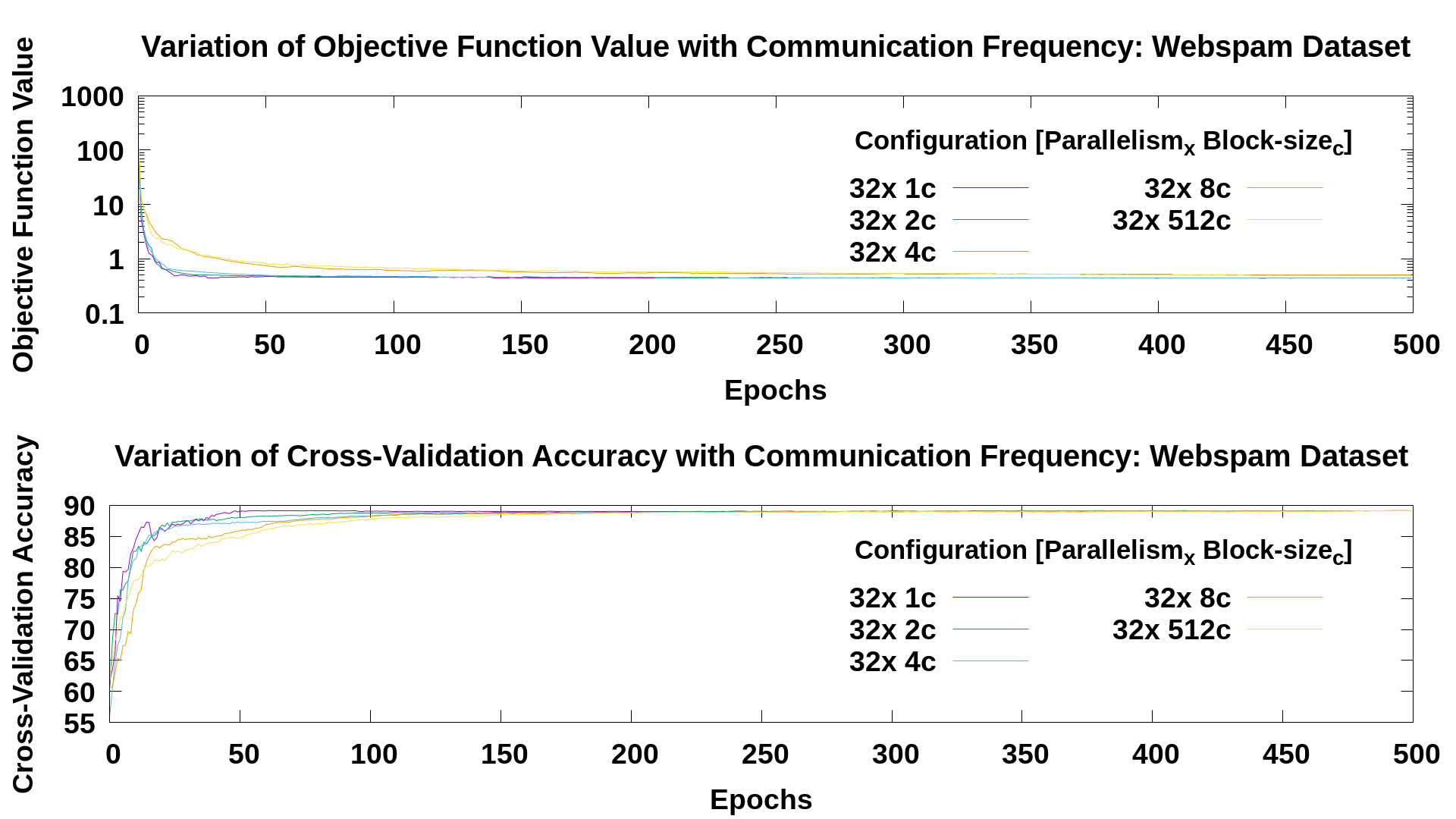}
\caption{Distributed Model Synchronizing Algorithm on Webspam Dataset for Parallelism [32] and Block Size = [1,2,4,8,512]}
\label{fig:dms-cost-webspam-32x-1,2,4,512}
\end{figure}



\begin{figure}[t]
\includegraphics[height=0.45\textwidth,width=0.50\textwidth]{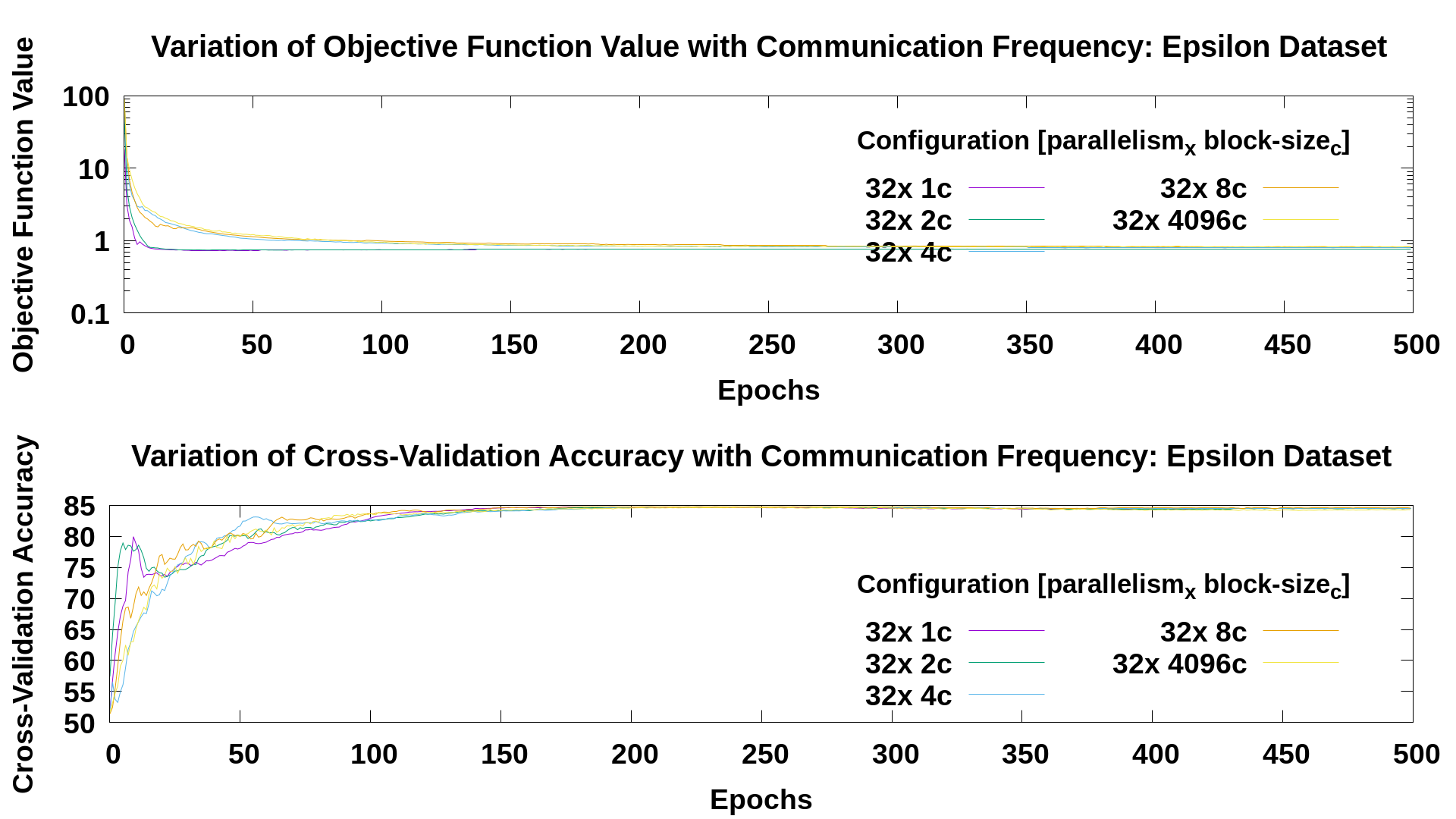}
\caption{Distributed Model Synchronizing Algorithm on Epsilon Dataset for Parallelism [32] and Block Size = [1,2,4,8,512]}
\label{fig:dms-cost-epsilon-32x-1,2,4,512}
\end{figure}

In summarizing the experiments done on distributed model synchronization on algorithm convergence, it is evident that the effect from the block size over convergence of the algorithm is approximately similar for all the block sizes explored in the experiments. 

\subsubsection{Distributed Model Synchronization on Efficient Training}

\begin{figure}[t]
\centering
\includegraphics[height=0.28\textwidth,width=0.50\textwidth]{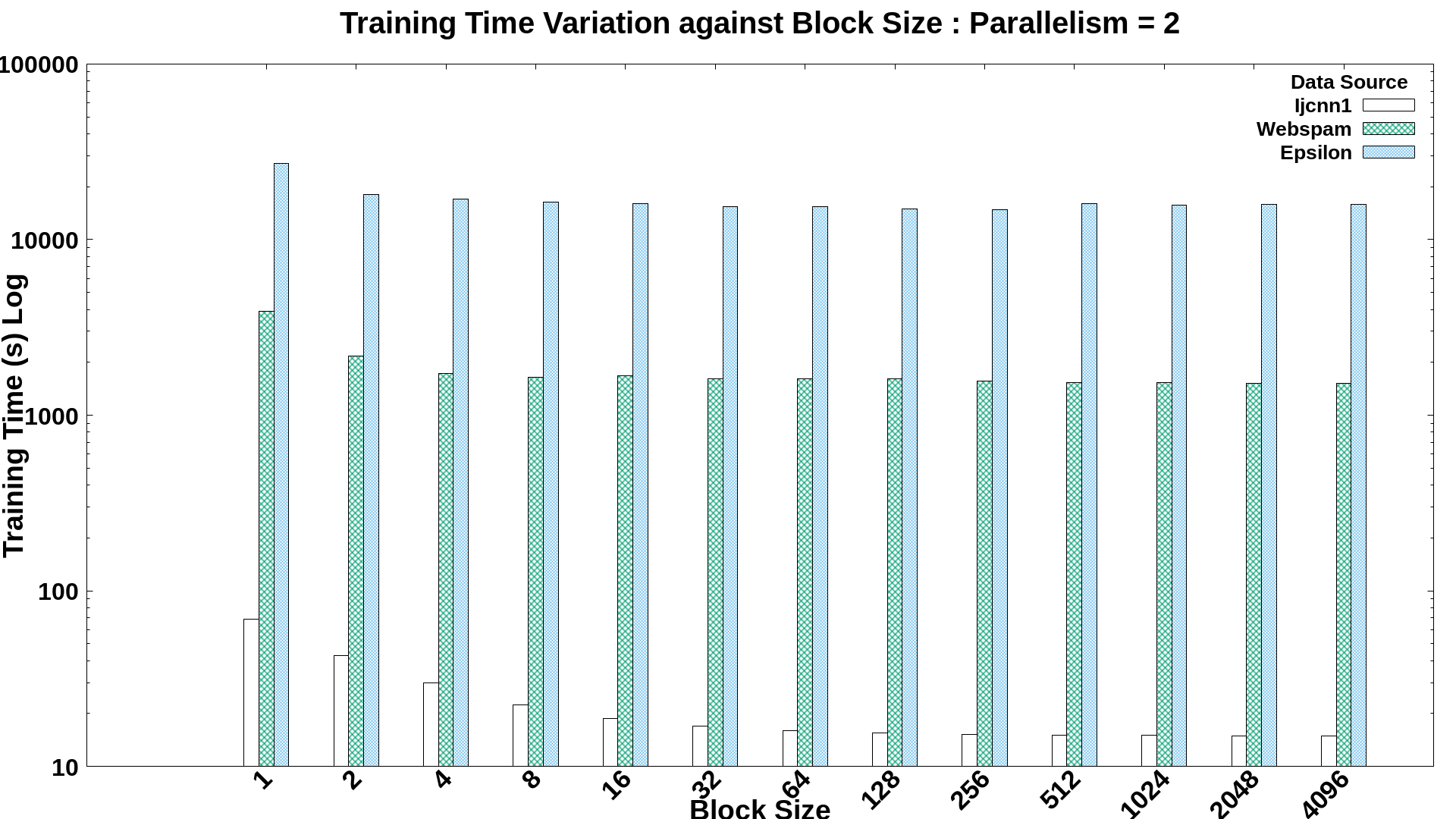}
\caption{Distributed Training Time Variation: Parallelism = 2, 80\% Training Set}
\label{fig:msf-at-all-log-2}
\end{figure}

\begin{figure}[t]
\centering
\includegraphics[height=0.28\textwidth,width=0.50\textwidth]{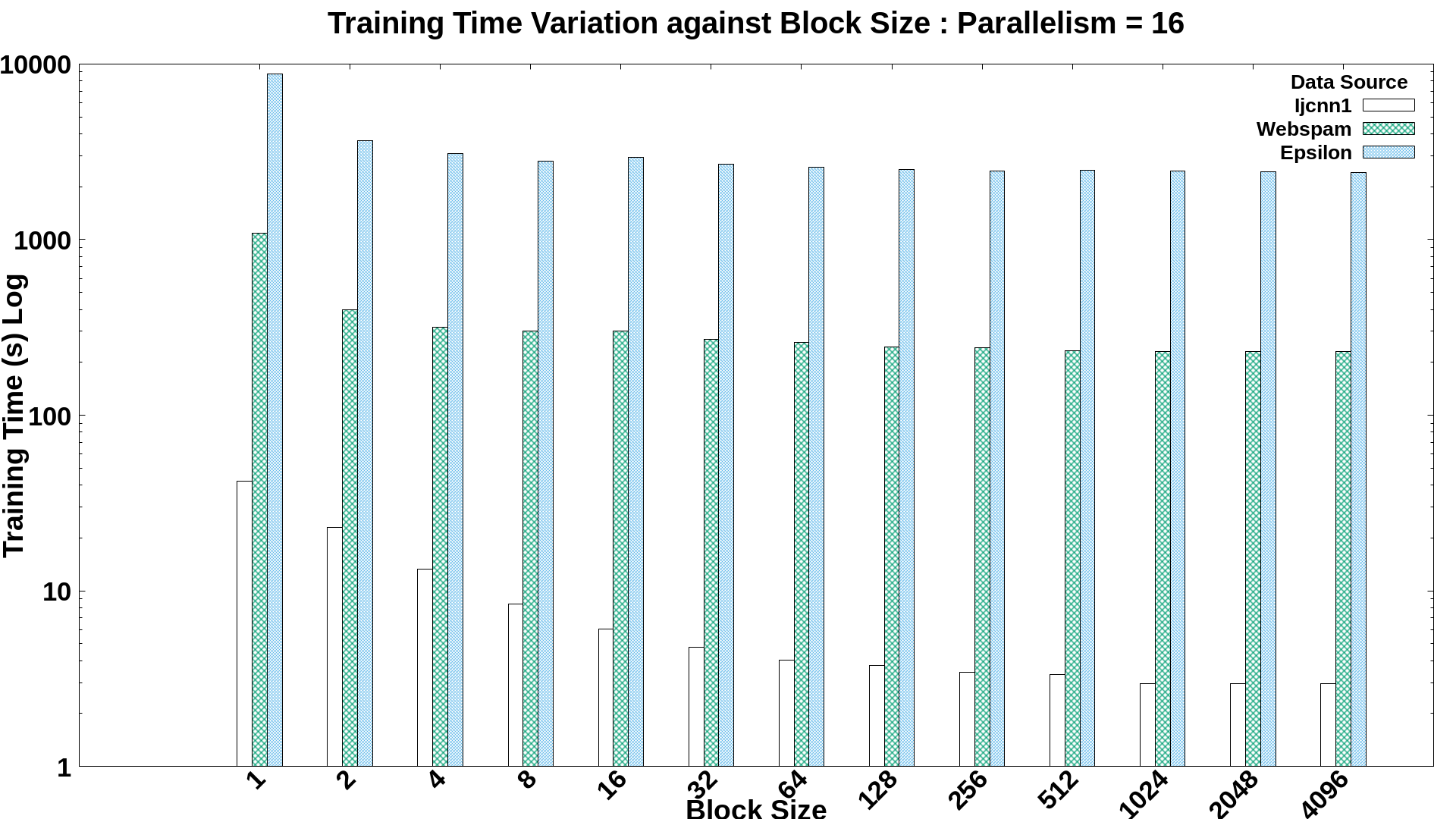}
\caption{Distributed Training Time Variation: Parallelism = 16, 80\% Training Set}
\label{fig:msf-at-all-log-16}
\end{figure}

\begin{figure}[t]
\centering
\includegraphics[height=0.28\textwidth,width=0.50\textwidth]{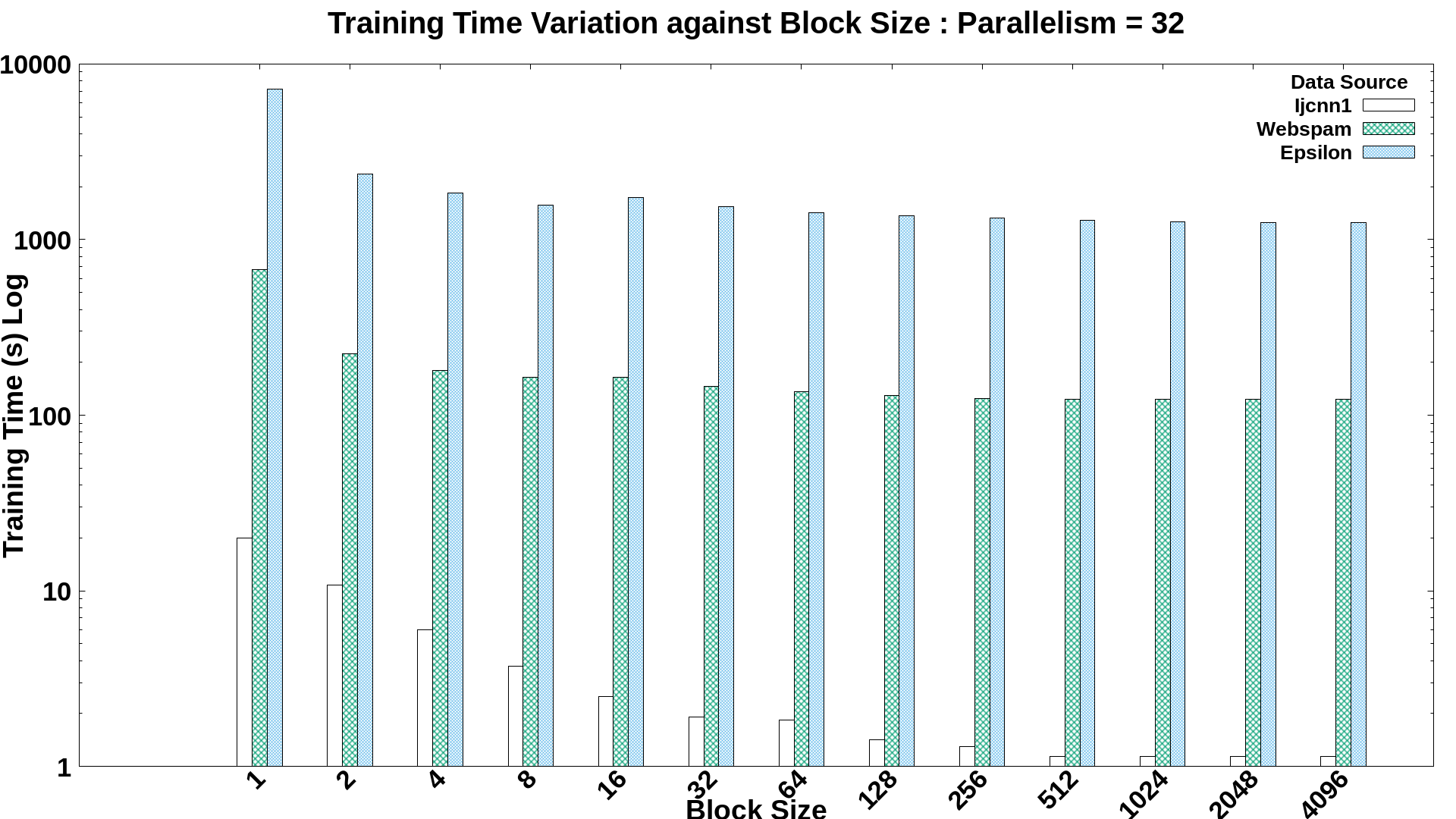}
\caption{Distributed Training Time Variation: Parallelism = 32, 80\% Training Set}
\label{fig:msf-at-all-log}
\end{figure}

Now we know that the MSF effect on the convergence of the algorithm is relatively independent on faster convergence. The next important factor is to determine how to select an MSF to obtain a fast training model. In analyzing this fact, first, we consider the overall training time needed on variable MSF with variable parallelisms. In figures \ref{fig:msf-at-all-log-2}, \ref{fig:msf-at-all-log-16} and \ref{fig:msf-at-all-log} the training time variation for variable MSFs on 2, 16 and 32 MPI processes is shown. It is evident from these results that higher MSF (lower block sizes) has a higher overhead in the training process. This effect dilutes down when the MSF is reduced. And the fluctuation diminishes at the lowest MSFs for each data set. In understanding the reason for the lagging in efficiency for higher MSF region, we conducted the same experiment by doing a benchmark on the communication and computation time for variable MSFs on variable parallelisms. 

\begin{figure*}[t]
\centering
   \caption{Training Time Breakdown in Data sets for Parallelisms 2, 80\% Training Set}
   \hbox to\linewidth{%
    \hfil%
    \vbox{%
        \hbox{\includegraphics[height=0.28\textwidth,width=0.30\textwidth]{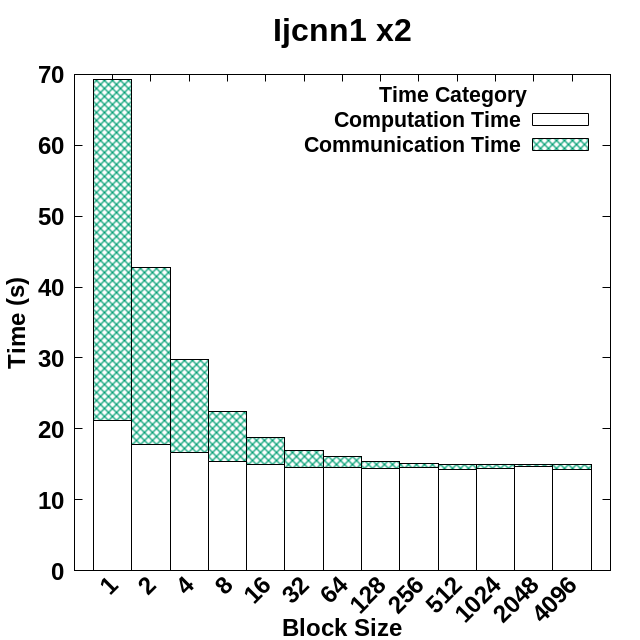}}%
    }%
    \hfil%
    \vbox{%
        \hbox{\includegraphics[height=0.28\textwidth,width=0.30\textwidth]{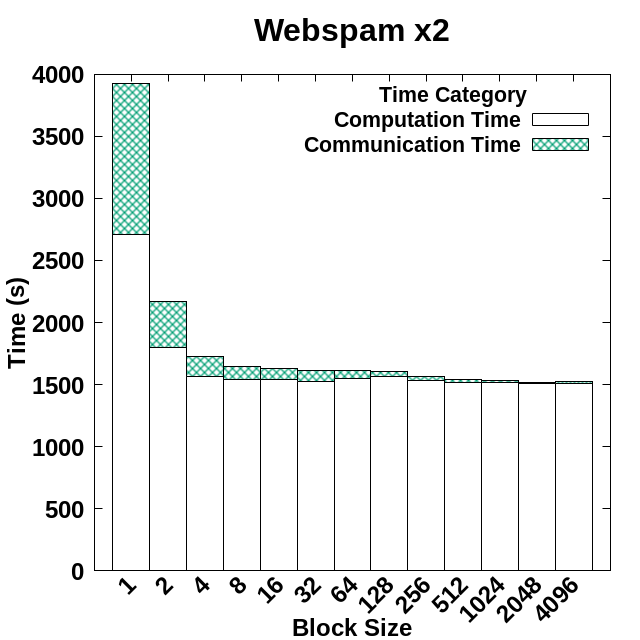}}%
    }%
    \hfil%
    \vbox{%
        \hbox{\includegraphics[height=0.28\textwidth,width=0.30\textwidth]{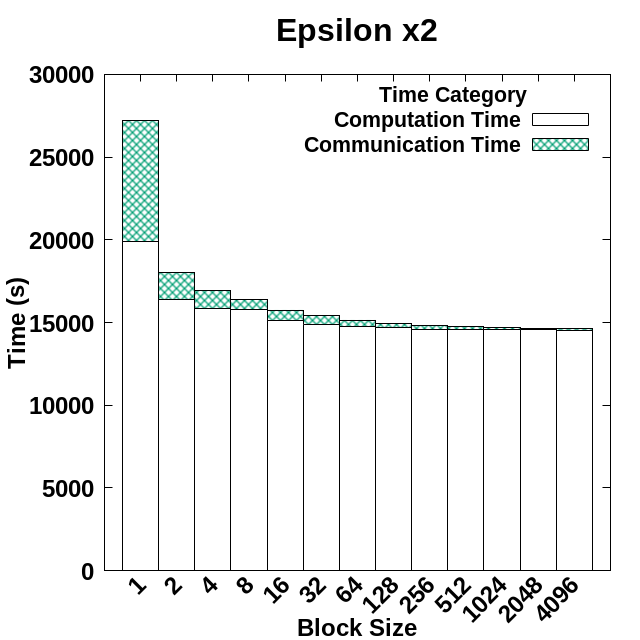}}%
    }%
    \hfil%
} 
\label{fig:msf-bt-all-log-2x}
\end{figure*}

\begin{figure*}[t]
\centering
   \caption{Training Time Breakdown in Data sets for Parallelisms 16, 80\% Training Set}
   \hbox to\linewidth{%
    \hfil%
    \vbox{%
        \hbox{\includegraphics[height=0.28\textwidth,width=0.30\textwidth]{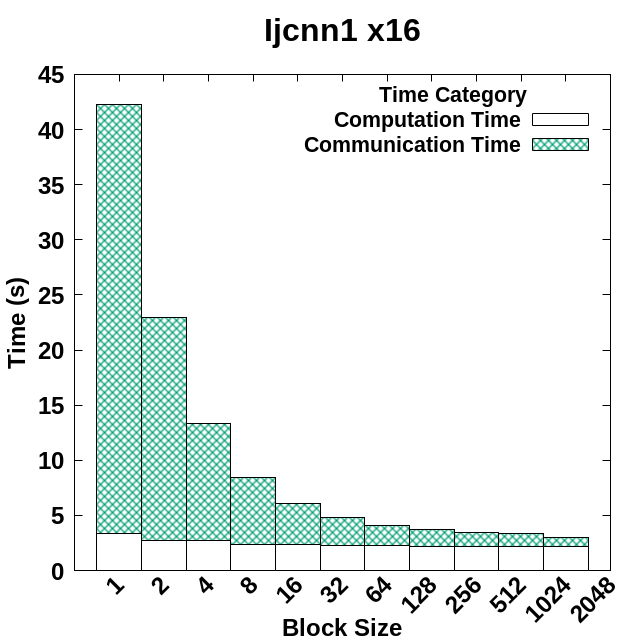}}%
    }%
    \hfil%
    \vbox{%
        \hbox{\includegraphics[height=0.28\textwidth,width=0.30\textwidth]{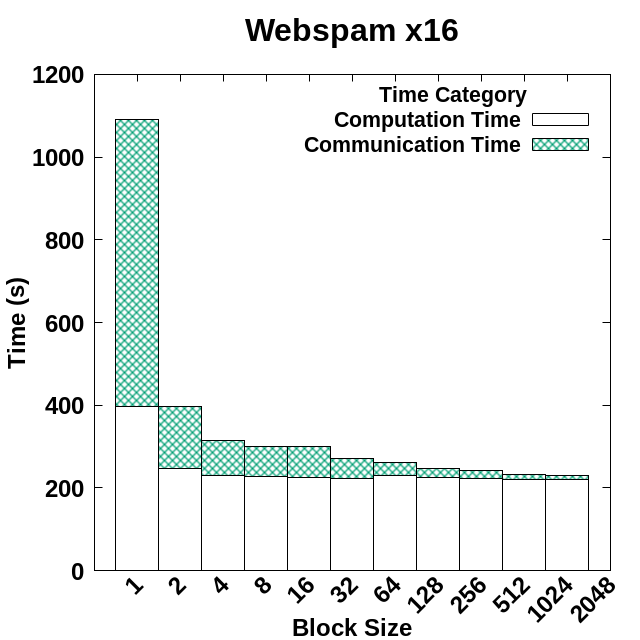}}%
    }%
    \hfil%
    \vbox{%
        \hbox{\includegraphics[height=0.28\textwidth,width=0.30\textwidth]{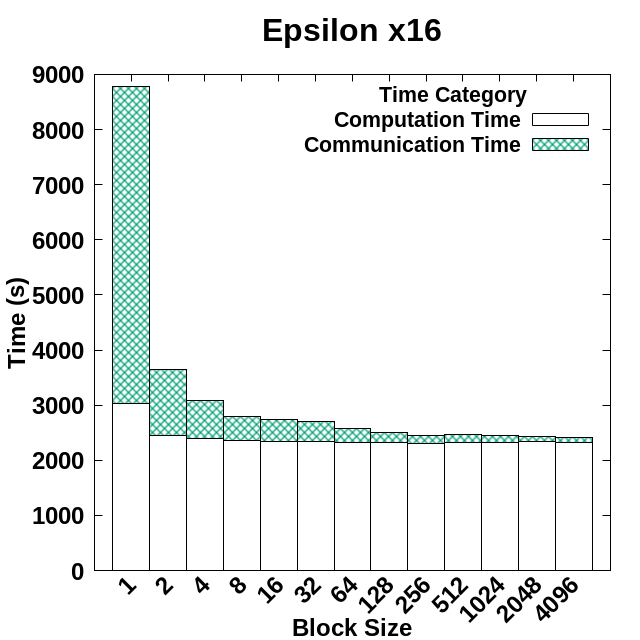}}%
    }%
    \hfil%
} 
\label{fig:msf-bt-all-log-16x}
\end{figure*}

\begin{figure*}[t]
\centering
   \caption{Training Time Breakdown in Data sets for Parallelisms 32, 80\% Training Set}
   \hbox to\linewidth{%
    \hfil%
    \vbox{%
        \hbox{\includegraphics[height=0.28\textwidth,width=0.30\textwidth]{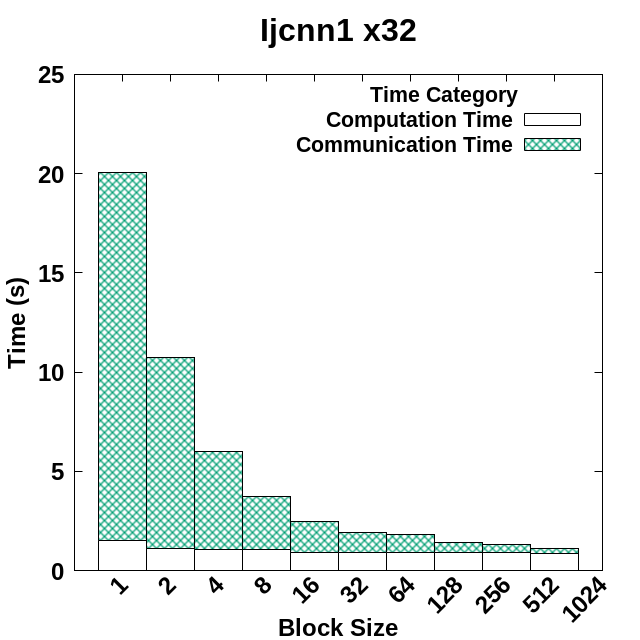}}%
    }%
    \hfil%
    \vbox{%
        \hbox{\includegraphics[height=0.28\textwidth,width=0.30\textwidth]{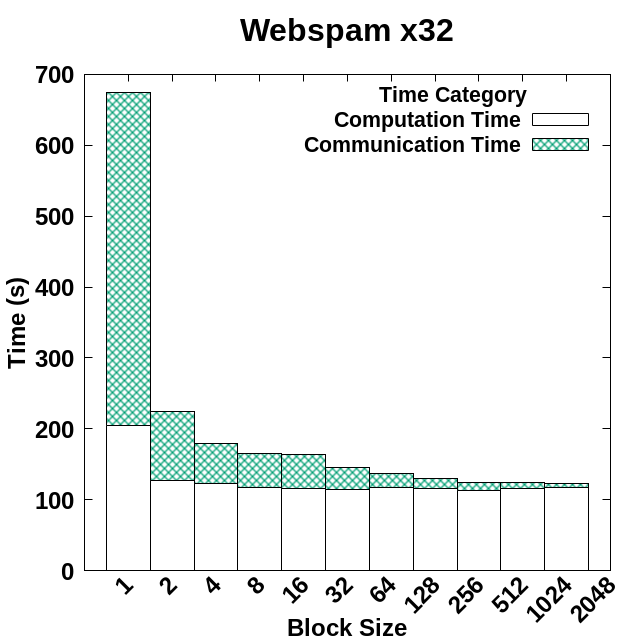}}%
    }%
    \hfil%
    \vbox{%
        \hbox{\includegraphics[height=0.28\textwidth,width=0.30\textwidth]{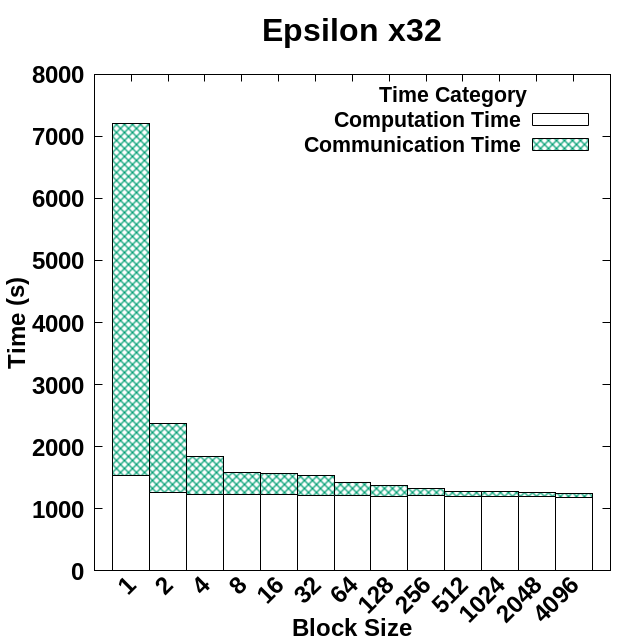}}%
    }%
    \hfil%
} 
\label{fig:msf-bt-all-log}
\end{figure*}

The training time breakdown for Ijcnn1, Webspam and Epsilon data sets for parallelism of 32, 16 and 2 are in figures \ref{fig:msf-bt-all-log}, \ref{fig:msf-bt-all-log-16x} and \ref{fig:msf-bt-all-log-2x}, respectively. In the higher MSF region (block sizes 1-8) the communication overhead is very high with respect to the lower MSF region (block size beyond 8). It is evident from this experiment that when MSF is higher the total communication overhead is high and it makes the training model run much slower than that of lower MSF. In considering the total communication overhead, there are two main factors governing the communication delay. The main reason is the MSF value and the second reason is the magnitude of the vector which is communicated over the network or over the processes. Ijcnn1 communicates a vector with 22 dimensions, Webspam a vector of 254 dimensions and Epsilon a vector of 2000 dimensions. It is clear from the time breakdown experiments that, when the vector shared over the network or processes has a higher dimension, the communication overhead is very high. When training data sets with higher data sizes and higher dimensions the communication overhead are very high. To mitigate this issue in training, we can increase the block size (decreasing MSF) so that a minimum amount of communication is done. Along with higher MSF, there is an additional computation overhead added in calculating the cross-validation accuracy and objective function value per each communication. The algorithm convergence is decided by the optimum value recorded by each of these variables. So each time we do a communication to synchronize the model, we check the convergence of the algorithm. This computation overhead is proportional to the size of the cross-validation data set. This effect dilutes down to a negligible value when a lower MSF is used. The effect can be clearly seen when the computation time for block size 1 and 2 are analyzed. The computation time for block size 1 is obviously high for all the data sets, but this effect dilutes down as the MSF lowers. But the effect from this computation overhead on the efficiency of the algorithm is lesser than that of communication overhead.

In this section, the conducted experiments elaborates that the high communication overhead is associated with the least block sizes and lesser communication overhead can be achieved by increasing the block size in communication. And also the ratio of computation time to communication time is very high in the lower block sizes. 

\subsubsection{Distributed Model Synchronization on Prediction Model}

\begin{figure}[t]
\centering
\includegraphics[height=0.28\textwidth,width=0.50\textwidth]{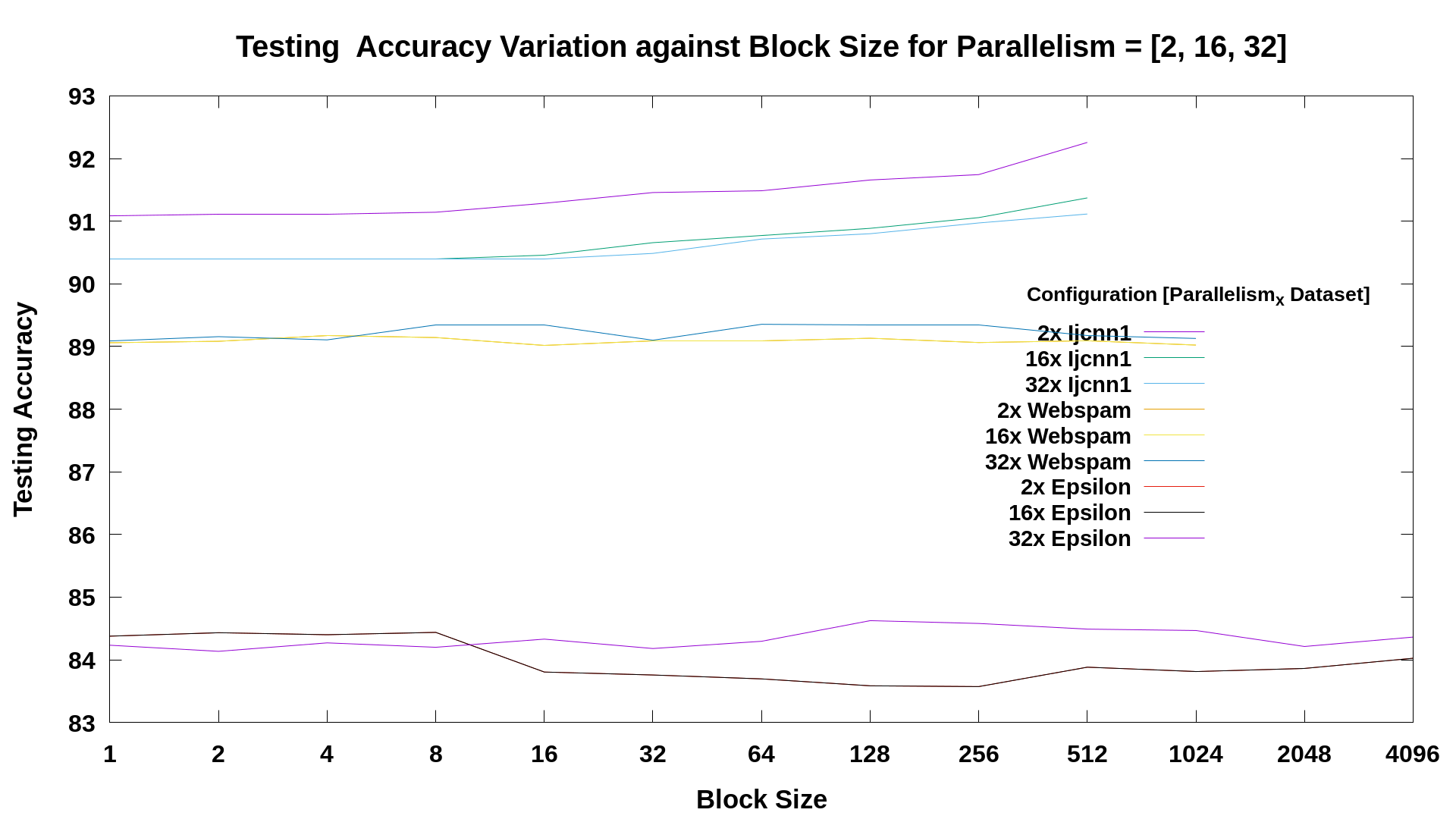}
\caption{Distributed Testing Accuracy Variation with Block Size: Parallelism 32}
\label{fig:msf-at-all-test-acc}
\end{figure}

The most important fact after training is to obtain a higher testing accuracy from the optimized trained model. It is important to see whether there is an effect from MSF on the testing accuracy for each data set. In figure \ref{fig:msf-at-all-test-acc}, the testing accuracy variation on variable MSFs for all three datasets is shown with respect to the parallelisms of 2, 16 and 32. It is evident from these results that the testing accuracy is not affected not more than $\pm 1.0\%$ in all three data sets. This observation provides us with the final proof we needed to understand the effect from the MSF variation over a highly optimized training model. The MSF effect on prediction model, convergence and efficiency of the training model shows that a lower MSF is the best fit to train the SGD-based SVM algorithm to obtain an efficient high accurate model. Table \ref{tb:summary} shows the summary of the experiments done in this research covering the aspects of improved efficiency, high testing accuracy and better speed up. From these results, it is clear that the MSF optimized distributed model training is a better solution in training larger data sets with SVM.

\begin{table}[t]
\caption{Experiment Summary : 60\% Training Data}
\begin{center}
\small\addtolength{\tabcolsep}{-4pt}
\begin{tabular}{|l|l|l|l|l|l|}
\hline
\multicolumn{1}{|c|}{\textbf{Dataset}} & \multicolumn{1}{c|}{\textbf{\begin{tabular}[c]{@{}c@{}}Sequential \\ Timing\\ (s)\end{tabular}}} & \multicolumn{1}{c|}{\textbf{\begin{tabular}[c]{@{}c@{}}Parallel \\ Timing\\  (s)\end{tabular}}} & \multicolumn{1}{c|}{\textbf{\begin{tabular}[c]{@{}c@{}}Sequential\\ Accuracy\end{tabular}}} & \multicolumn{1}{c|}{\textbf{\begin{tabular}[c]{@{}c@{}}Parallel\\ Accuracy\end{tabular}}} & \multicolumn{1}{c|}{\textbf{\begin{tabular}[c]{@{}c@{}}Speed\\  Up\\ (x1 vs x32)\end{tabular}}} \\ \hline
Ijcnn1 & 22.19 & 1.37 & 90.63 & 91.51 & 16.20 \\ \hline
Webspam & 2946.49 & 120.02 & 87.69 & 89.12 & 24.55 \\ \hline
Epsilon & 22208.07 & 968.782 & 80.06  & 84.36 & 22.92 \\ \hline
\end{tabular}
\label{tb:summary}
\end{center}
\end{table}

\section{Conclusion and Future Work}\label{s:conclusion}

In understanding the distributed scaling of the SGD-based SVM algorithm, it is vital to keep track on the effect of the model synchronization frequency on the convergence of the algorithm. The convergence is governed by the cross-validation accuracy and value of the objective function. Both parameters need to be simultaneously optimized such that the fluctuation of each variable must be the least or satisfy an expected threshold where we call, the algorithm has reached the convergence. The final outcome from the trained model is to guarantee the expected high accuracy in the testing phase. In our research, it is evident that the model synchronization frequency is a key factor as far as fast execution and fast convergence are considered. 

Our research also shows that high MSF value is loosely coupled with the convergence of the training model, test accuracy and efficient execution. This allows us to use the MSF as lower as possible to obtain a fast training model with high accuracy. The efficiency of the training model is highly valuable when training large data sets with higher dimensions and this is evident with the observations we made on Ijcnn1, Webspam and Epsilon data sets. With reference to the standard SGD-based sequential SVM algorithm, our distributed model provides higher accuracy for all three data sets with less over-fitting on training data. This shows that our model synchronization optimization is better in gaining a higher testing accuracy with efficient execution. 

We are currently working on implementing this idea with our big data toolkit Twister2 for both batch and streaming training of SVM. Our streaming idea is still experimental and we are working on improving the streaming model based on the findings of this research. Model update of a system in production is another aspect we are evaluating with the findings from this research. 

Towards developing streaming and batch based machine learning applications with edge devices, it is vital to obtain maximum performance from resources. Our final objective is to expand this model on training high volume data in edge devices in sensory networks to obtain highly efficient training with limited resources.

\section*{Acknowledgment}

This work was partially supported by NSF CIF21 DIBBS 1443054. We extend our gratitude to NSF for providing us funding to continue this project. And also we extend our gratitude to Future Systems team for their support with infrastructure to run our system and Digital Science Center for their guidance and the support to complete this research.






\vspace{12pt}

\end{document}